\documentclass[aps, pra,preprint,showpacs,preprintnumbers,amsmath,amssymb]{revtex4-1}
\usepackage{amsmath}
\usepackage{latexsym}
\usepackage{txfonts}
\usepackage{graphicx}
\usepackage{amssymb}
\usepackage{float}
\usepackage{subcaption}

\newcommand {\beq}{\begin {equation}}
\newcommand {\eeq}{\end {equation}}

\newcommand {\ptq}{\frac{\partial^{2}}{\partial t^{2}}}
\newcommand {\pzq}{\frac{\partial^{2}}{\partial z^{2}}}
\newcommand {\pt}{\frac{\partial}{\partial t}}
\newcommand {\pz}{\frac{\partial}{\partial z}}
\newcommand {\ptprime}{\frac{\partial}{\partial t^{\prime}}}

\newcommand {\pzeta}{\frac{\partial}{\partial \zeta}}

\newcommand {\cq}{c^2}

\newcommand {\tprime}{t^{\prime}}

\newcommand {\vvperp}{{\vec{v}_\perp}}

\begin{document}


\title{Coherent $\pi$-pulse emitted by a dense relativistic cold electron beam }
\author{J. A. Arteaga$^1$, L. F. Monteiro$^1$, A. Serbeto$^1$, K. H. Tsui$^1$, J. T. Mendon\c{c}a$^2$}
\affiliation{$^1$Instituto de F\'isica, Universidade Federal Fluminense, Campus da Praia Vermelha, Niter\'oi, RJ,  24210-346, Brasil,}
\affiliation{$^2$IPFN, Instituto Superior T\'ecnico, Universidade de Lisboa, 1049-001 Lisboa, Portugal}
\email{johny@if.uff.br, lfm@if.uff.br,  serbeto@if.uff.br, tsui@if.uff.br, 	titomend@ist.utl.pt} 

\begin{abstract}

Starting from a three-wave interaction system of equations for free-electron lasers  in the framework of a quantum fluid model, we show that these equations  satisfy the Sine-Gordon equation. The full solution in space and in time of this set of equations are numerically obtained. 
\end{abstract}

\maketitle


Devices as Free-Electron Lasers (FELs) constitute a very active field of research nowadays. It can generate high-power coherent radiation by releasing the kinetic energy of a relativistic electron beam into field energy, whose  main caracteristics are the tunneability and the possibility of reaching extremely small wavelengths, such X-ray and $\gamma$ rays, not acessible to molecular lasers.  They were conceived in the early seventies by using Quantum Mechanics approach\cite{madey1971}. It was shown subsequently that models based on classical electromagnetism works as well, if the emited photon momentum recoil is not greater than the electron momentum spread\cite{hopfclassical,tkwanclassical,krollclassical}.

In a classical point of view, electromagnetic radiation can be generated by forcing a relativistic electron beam to oscillate inside a wiggler, or undulator, which consists of a suitable arrangement of magnets with alternating poles that produces a magnetic field of period $\lambda_w$.  As the beam travels through
 the device,  the electrons are subject to a Lorentz force and emit in his own reference frame a spontaneous Lorentz-contracted radiation of wavelength $\lambda_{w}^{\prime} = \lambda_w/\gamma$, where $\gamma$ is the beam relativistic factor.  Back to the laboratory frame, this radiation is again Lorentz contracted, resulting in an output radiation  with a much shorter wavelength given by $\lambda_s = \lambda_w/2\gamma^2$, where the subscript (s) stands for scattered radiation.
However, this so called undulator radiation is still incoherent, not a laser. If the wiggler is long enough and the current is high, the radiation interacts back with the beam, which forces the electrons to group into microbunches in the scale of the radiation wavelength.  As they are almost in the same phase, they emit collective coherent radiation \cite{boniilnuovocimento}, which can operate in a frequency range not accessible to conventional molecular lasers, such as extreme ultraviolet and X-Ray radiation, by just adjusting the wiggler period parameter and the beam energy. This is the essence of a Free-Electron Laser in a Self Amplified Spontaneous Emission (SASE) regime, which is the heart of the biggest X-Ray FEL facilities nowadays, as for example DESY in Germany, SLAC X-Ray Laser, in United States, and SACLA, in Japan.
However, for this angstron-scale wavelengths, the classical SASE has a drawback. Since an electron beam bunch usually encompasses a large number of cooperation lengths and considering that each cooperation length emits randomly from each other, the result is a chaotic spiking behaviour and a poor coherent spectrum to the laser output.  The number of spikes corresponds approximately to $L_b / 2\pi L_c$, where $L_b$ is the length of the electron bunch, $L_c = \lambda_s/ 4 \pi \rho$ is the cooperation length, and $\rho$ is the classical FEL parameter, defined by $\rho = \gamma^{-1}(a_w/4c k_w)^{2/3}(e^2 n_b/m_e \epsilon_0)$, where $a_w$ is the wiggler parameter and $n_b$ is the unperturbed beam density\cite{Boni1994_73_PRL, Boni2005b}.
A way to overcome such a problem is the FEL to operate in the so called quantum regime.  In this regime, the electron emits only a single photon and a phenomenon of ``quantum purification'' occurs, which means that the noisy spectrum of the classical SASE reduces to a single narrow line, which is limited by the electron beam duration, improving largely the coherence of the output radiation\cite{Boni2008}.
In order to realize a Quantum FEL, it is more feasible to use a counterpropagating optical wiggler instead of a magnetostatic one \cite{Boni2005d}.  Besides the advantage of ``quantum purification'', the resonance dispersion relation for an electromagnetic wiggler turns into approximately $\lambda_s = \lambda_w/4\gamma^2$, and $\lambda_w$ can now be submilimetric, instead of the order of centimeter scale.  This reduces the beam energy requirements and consequently the length of the electron accelerator, opening a window to table top X-Ray FELs, avoiding multi GeV accelerators to achieve the same frequency output.

In previous works the production of visible laser radiation during the interaction of an electromagnetic pump with a low energy electron beam was analysed by Lin and Dawson\cite{dawson1980} by using classical electromagnetism. Mendon\c{c}a and Serbeto\cite{mendonca}, using the relativistic electron fluid model, showed that ultrashort gamma ray might be obtained by a collective scattering of an incident infrared laser radiation on energetic electron bunches, that works as imperfect relativistic mirrors. Serbeto \emph{et. al.} \cite{serbetoqfel} and L. F. Monteiro \emph{et. al} \cite{lfmonteiro}, by using a quantum hydrodynamic model, presented a set of nonlinear quantum plasma fluid equations to describe the quantum free-electron laser.
In this model a set of nonlinear coupled equations is obtained, where the optical wiggler, the beam plasma mode, and the scattered radiation(seed) interact as a three-wave process.  Here, we extend the previous numerical results to a full solution in space and in time for such system of 
equations by using a more realistic gaussian profile for the seed radiation.


In previous works L. F. Monteiro \emph{et. al} presented a quantum fluid model for a cold electron beam interacting with an electromagnetic field. It was shown that the evolution of the space-chage oscillation is given by \cite{lfmonteiro}

\begin{equation}
\left[\frac{\partial^2}{\partial {t^{\prime}}^{2}} + \left(\frac{\hbar}{2 m_{e} \gamma_{e}^3}\right)^2 \frac{\partial ^4}{\partial \zeta^4}  +  \frac{\omega_{p}^2}{\gamma_{e}^3}\right] \frac{\delta n}{n_b}
 = \frac{1}{m_e \gamma_{e}^3}\frac{\partial^2}{\partial \zeta^2} V,  \label{plasmaoscilation}
\end{equation}
where $V$ is the ponderomotive potential, which arises from the beating of a pump(optical wiggler) and a scattered radiation(seed), $\omega_{p} = (4\pi e^2 n_b/m_e)^{1/2}$ is plasma frequency, $n_b$ is the unperturbed beam density, $\delta n /n_b$ is the normalized space-charge oscilation amplitude, $\gamma_e$ is the normalized energy associated to the unperturbed longitudinal beam velocity $v_e$, and $m_e$ is the electron rest mass.   It is worth emphasizing that this equations is deduced in a $\zeta$ frame  that is moving with the electron beam, and has exactly the same form as the forced quantum plasma wave equation in the laboratory frame. 
In order to get a self-consistent scheme, we need to obtain an expression for the potential $V$ as a function of the electromagnetic fields.  First of all, let us take the wave equation, viz.,

\begin{equation}
\left( \pzq - \frac{1}{\cq}\ptq \right) (\vec{A}_s + \vec{A}_w) = - \frac{4 \pi}{c} \vec{J}_{\perp}, \label{waveequation}
\end{equation}
where $\vec{J}_{\perp} = -|e| n \vvperp$ is the perpendicular current. Due to transverse momentum conservation, in the laboratory frame, the perpendicular velocity is given by
\begin{equation}
\vvperp =  \frac{|e|(\vec{A}_{s} + \vec{A}_{w})}{\gamma m_e c} \label{vperp}
\end{equation}
 where $\gamma$ is the electron relativistic factor. Here, $\vec{A}_{s}$ and $\vec{A}_{w}$ stand for the potential vectors of the seed and optical pump fields, which are assumed circularly polarized waves. Considering that the seed radiation propagates in the negative $\hat{z}$ direction (the same as the electron beam) and the pump propagates in the counter direction, these fields can be represented as

\begin{equation}
\vec{A}_{s}(z,t) = -\frac{1}{\sqrt{2}} \left(\hat{e}i A_s e^{i(k_{s} z + \omega_{s}t)} + cc \right), \label{AS}
\end{equation}

\begin{equation}
\vec{A}_{w}(z,t) = \frac{1}{\sqrt{2}}\left(\hat{e}A_w e^{i(-k_{w}z + \omega_{w} t)} + cc \right), \label{AW}
\end{equation}
where $\hat{e} = (\hat{x} + i\hat{y})/\sqrt{2}$ is the unitary polarization vector and has the properties $\hat{e}\cdot\hat{e} = \hat{e}^{*}\cdot\hat{e}^{*}= 0$, $\hat{e}\cdot\hat{e}^{*} = \hat{e}^{*}\cdot\hat{e}=1$ . Here, $A_{s}(A_{w})$ is the envelope amplitude with $k_{s}(k_{w})$ and $\omega_{s}(\omega_{w})$ being the wavenumber and frequency of the scattered radiation(optical wiggler).
The ponderomotive force due to these fields, which acts on the electron beam, can be defined as
\begin{equation}
\vec{F}_p = -\nabla V = -m_e \cq \nabla \gamma.
\end{equation}
Here, $\gamma = \sqrt{1 + p_{e}^{2}/m_{e}^{2}c^{2} + p_{\perp}^{2}/m_{e}^{2}c^{2} }$ is the electron  relativistic factor, that can be written as $\gamma \approx \gamma_e$ in the denominator if $p_{\perp}^{2}/m_{e}^{2}c^{2} = {a_t}^2 << 1$, where $\vec{a_{t}} = \vec{a_{s}} + \vec{a_{w}}$ is the total normalized potential vector. Assuming $(|a_{w}|^{2} + |a_{s}|^{2})/\gamma_{e}^{2} << 1$ and neglecting nonresonant terms, we obtain the corresponding ponderomotive potential,
\begin{equation}
V = i\frac{mc^2}{2 \gamma_e}\left(a_{w} a_{s}^{*} e^{i\theta_{p}} - cc \right),  \label{pondpot}
\end{equation}
with $a_{s} = |e| A_{s}/m_{e} c^2$($a_{w} = |e| A_{w}/m_{e} c^2$) being  the normalized radiation(optical wiggler) wave amplitude, and $\theta_p = -(k_{s} + k_{w})z - (\omega_{s} - \omega_{w})t$ is the ponderomotive phase in the laboratory frame. 

Assuming that the beam is travelling to the left with velocity $-v_e \hat{z}$, we can write Eq. (\ref{plasmaoscilation}) from the point of view of the laboratory frame by doing the inverse transformations $z = \zeta - v_e \tprime$ and $t = \tprime$. So, the derivatives transform according to
\beq
\ptprime =  \pt - v_e \pzeta; \   \   \   \   \pzeta = \pz.
\eeq

By using the expression for the ponderomotive potential defined by Eq.(\ref{pondpot}) and considering that the radiation perturbed envelope has slow variation in space and in time, we can, therefore, neglect the higher order derivatives. Assuming that the plasmon has the shape $\delta n /n_b = n(z,t) \textnormal{exp}(i(-k_l z- \omega_l t))$, where the amplitude $n(z,t)$ is also slow-varying and keeping valid the FEL matching conditions, $k_l = k_s + k_w$ and $\omega_w = \omega_s - \omega_l$, we obtain from Eq. (\ref{plasmaoscilation}),

\begin{equation}
\left[ \frac{\partial}{\partial t} - \left(\frac{1}{2\Omega_l}\left(\frac{\hbar}{m_e \gamma_e^3}\right)^2k_l^3  + v_e \right) \frac{\partial}{\partial z}\right] n =  \frac{c^2 k_l^2}{4\gamma_e^4 \Omega_l}  \, a_w \, a_s^*  \label{plasmaoscilationsv}
\end{equation}
where $\Omega_l = \omega_l -k_l v_e$ is the Doppler-shifted beam wave frequency.  Besides the matching conditions, in order to obtain Eq. (\ref{plasmaoscilationsv}) from Eq. (\ref{plasmaoscilation}) it is also assumed the following quantum corrected linear dispersion relation, viz.,
\begin{equation}
 \Omega_{l}^2 - \frac{w_{p}^2}{\gamma_{e}^3} - \frac{c^2 \lambdabar_{c}^2}{4 \gamma_{e}^6} k_{l}^4 = 0. \label{lineardispersionrelation}
\end{equation}

From the wave equation (\ref{waveequation}) and using the electromagnetic fields defined according to Eqs. (\ref{AS}) and (\ref{AW}), we obtain the following equations for the slow-varying envelope radiation described in the stationary frame of the laboratory, viz., 

\begin{equation}
\left[ \frac{\partial}{\partial t} - v_s \frac{\partial}{\partial z} \right] a_s = -\frac{\omega_p^2}{2 \gamma_e \omega_s} n^* a_w, \label{seedsv}
\end{equation}
\begin{equation}
\left[\frac{\partial }{\partial t} + v_w \frac{\partial}{\partial z} \right] a_w =
 \frac{\omega_p^2}{2 \gamma_e \omega_w} n a_s,  \label{pumpsv}
\end{equation}
where we have taken into account the linear plasma dispersion relation $\omega_{s(w)}^2 = \omega_{p}^2/\gamma_e + c^2 k_{s(w)}^2$ and  $v_{s,w} \equiv c^2 k_{s,w}/\omega_{s,w} $ is the radiation(optical wiggler) group velociy. Equations (\ref{plasmaoscilationsv}) and (\ref{seedsv})-(\ref{pumpsv}) form a set of coupled nonlinear fluid equations which describe the quantum FEL dynamics pumped by an electromagnetic wave.
These equations have already been solved numerically in time or in the steady-state regime in earlier articles.  Here,  we propose to attempt  solutions in the stationary frame of the scattered radiation through the following characteristic transformation

\begin{equation}
\chi^\dag = z + v_s t, \,\,\, \tau^\dag = t. \nonumber
\end{equation}
The space and time varibles are normalized such that $\chi = (\omega_p/c )\chi^\dag$ and $\tau = \omega_p \tau^\dag $, respectively. Hence,  the final set of equations to be solved can be rewritten as
\begin{eqnarray}
\frac{\partial a_s}{\partial \tau}  &=& - \frac{\omega_p}{2 \gamma_e \omega_s} \, a_w \, n^* = -C_s \, a_w \, n^* ,  \label{as} \\
\left( \frac{\partial}{\partial \tau} + (\beta_s + \beta_w) \frac{\partial}{\partial \chi}  \right) a_w &=& \frac{\omega_p}{2 \gamma_e \omega_w}\, a_s \, n   = C_w \, a_s \, n,  \label{aw}  \\
\left( \frac{\partial}{\partial \tau} + (\beta_s - \beta_e) \frac{\partial}{\partial \chi}  - \beta_q \frac{\partial}{\partial \chi} \right) n &=& -\frac{c^2 k_l^2}{4\gamma_e^4 \omega_p |\Omega_l|} \, a_w \, a_s^* = - C_l \, a_w \, a_s^* . \label{n}
\end{eqnarray} 
Here, $\beta_{s,w} = v_{s,w}/c$ and $\beta_q = \hbar^2 k_l^3/ 2\Omega_lm_e^2\gamma_e^6$. 

\begin{figure}[h!]

  \centering
  \includegraphics[width=0.7\linewidth]{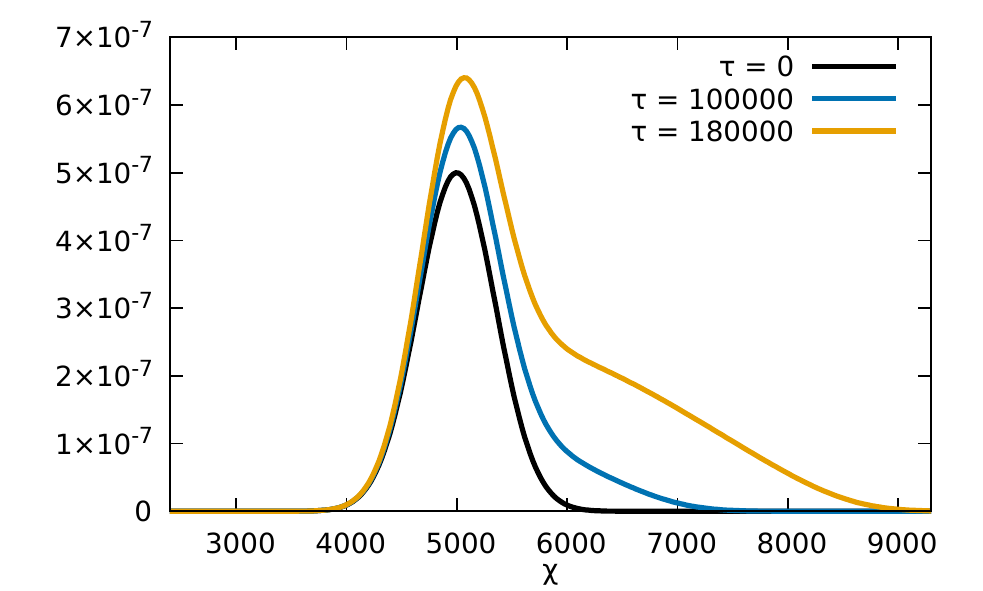}
  \caption{\footnotesize Spatial
   evolution of the seed radiation $a_s$ for three times of interaction. The initial  wiggler amplitude $a_{w}(\chi,0)= 1.0\times10^{-5}$, initial Gaussian seed profile of $a_s(\chi,0) = 5.0\times10^{-7}$, which corresponds to $5\%$ of the optical wiggler amplitude,  centered at $\chi_0 = 5000$ and a half-width of $\Delta \chi = 500$.}.   
  \label{fig:linear} 
  \normalsize
\end{figure}

Equations (\ref{as})-(\ref{n}) are solved numerically in space and in time taking into account an electron beam with normalized energy, $\gamma_{e} = 5.0$, and density, $n_{b} = 5.0\times 10^{18} cm^{-3}$, interacting with optical wiggler
with an initial amplitude, $a_{w} = 1.0\times10^{-5}$.  As initial conditions we have assumed a  Gaussian pulse profile, $a_s = a_{s}(\chi,0) e^{\left(\frac{\chi - \chi_0}{\Delta \chi}\right)^2}$, for the seed radiation. A constant shape pump, with a longitudinal size larger than radiation, is considered, and a null density perturbation. In Figure \ref{fig:linear} we present the seed evolution at the beginning of the interaction, when $a_w$ still remains constant.  
We observe the gaussian pulse amplification followed by an asymmetrical increase in half-width pulse, which is due to the different velocities of propagation inside it \cite{Bobroff} and the pulse undergoes a change in its shape before entering the stage of optical wiggler depletion \cite{trines2, lehman}. 

\begin{figure}[h!]
  \includegraphics[width=2.8in]{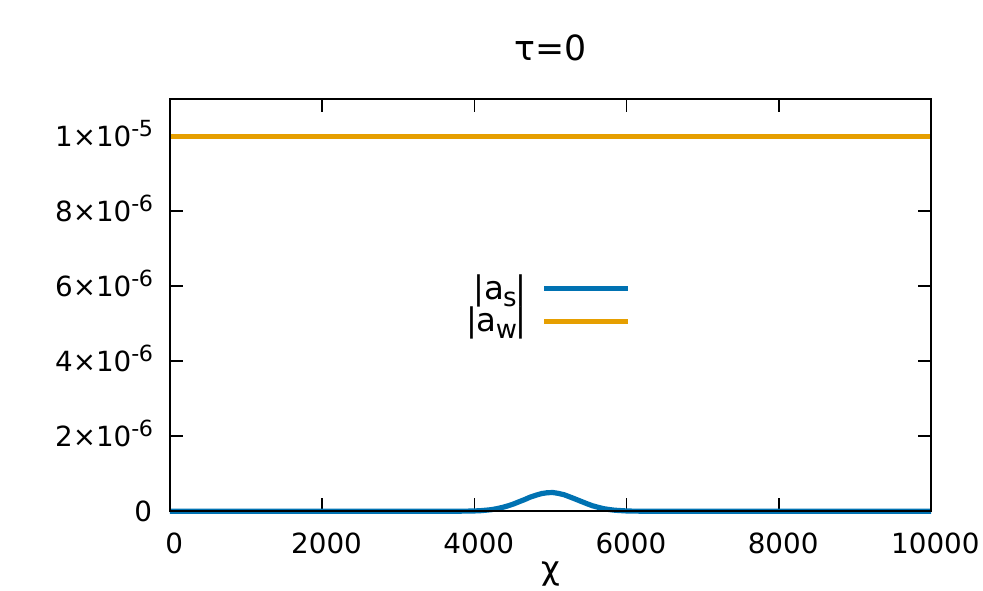}
  \includegraphics[width=2.8in]{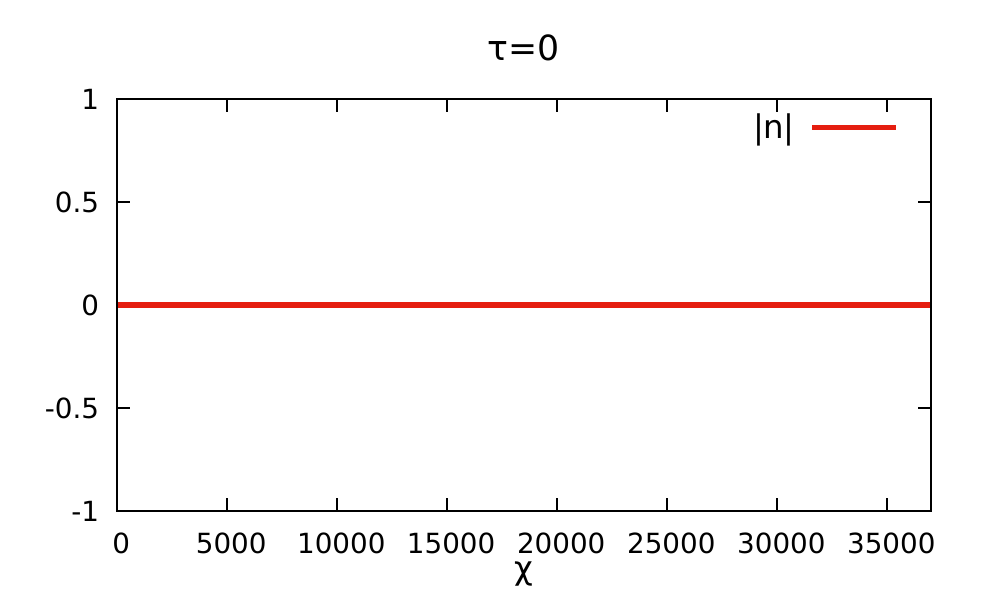}
  \includegraphics[width=2.8in]{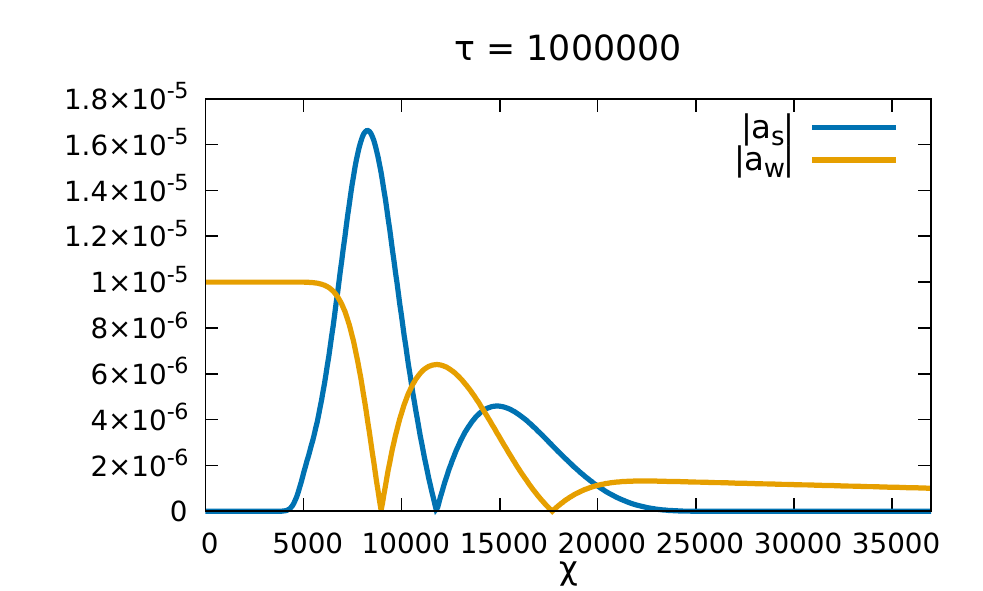}
  \includegraphics[width=2.8in]{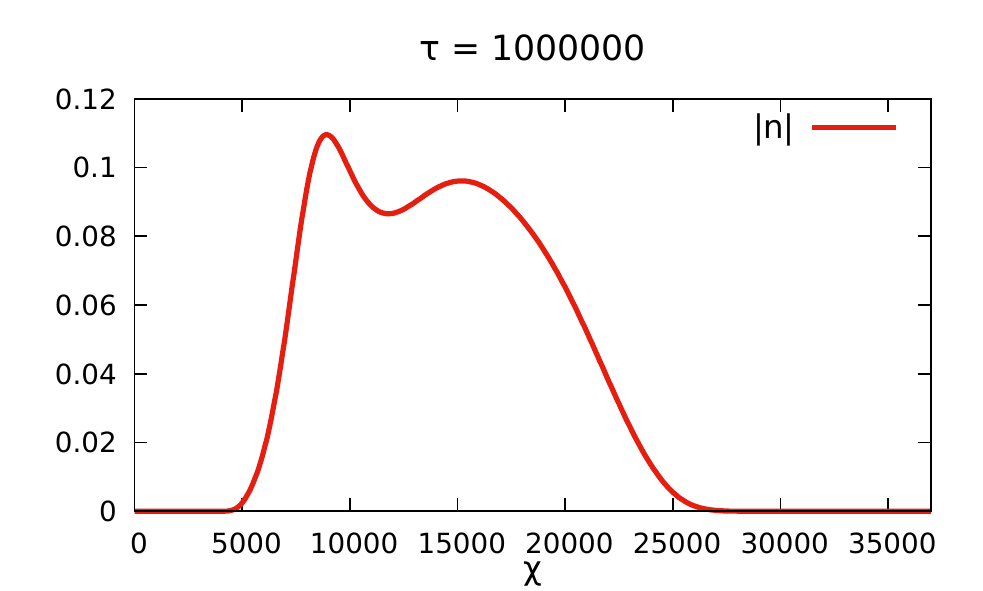}
  \includegraphics[width=2.8in]{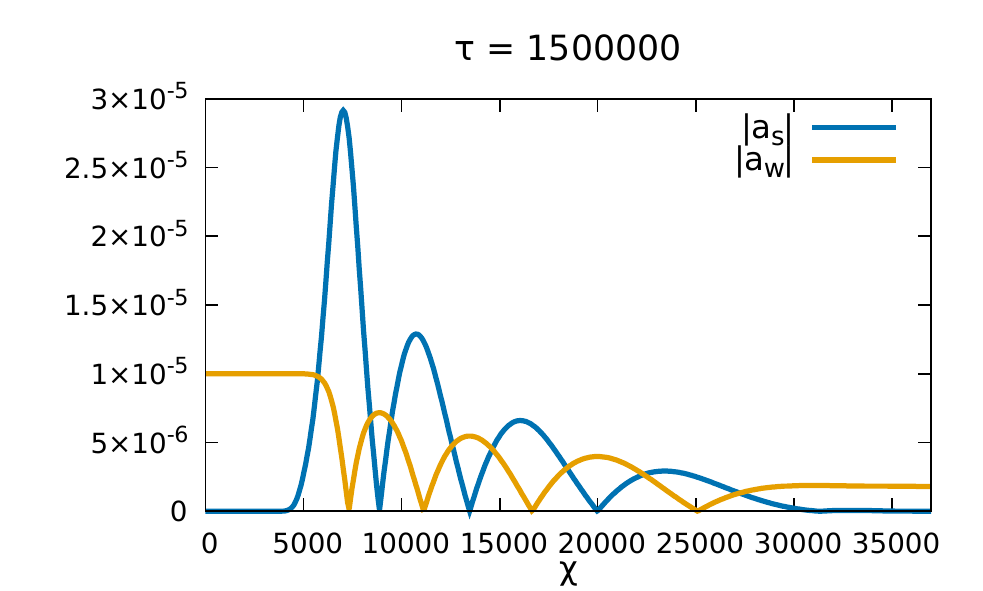}
  \includegraphics[width=2.8in]{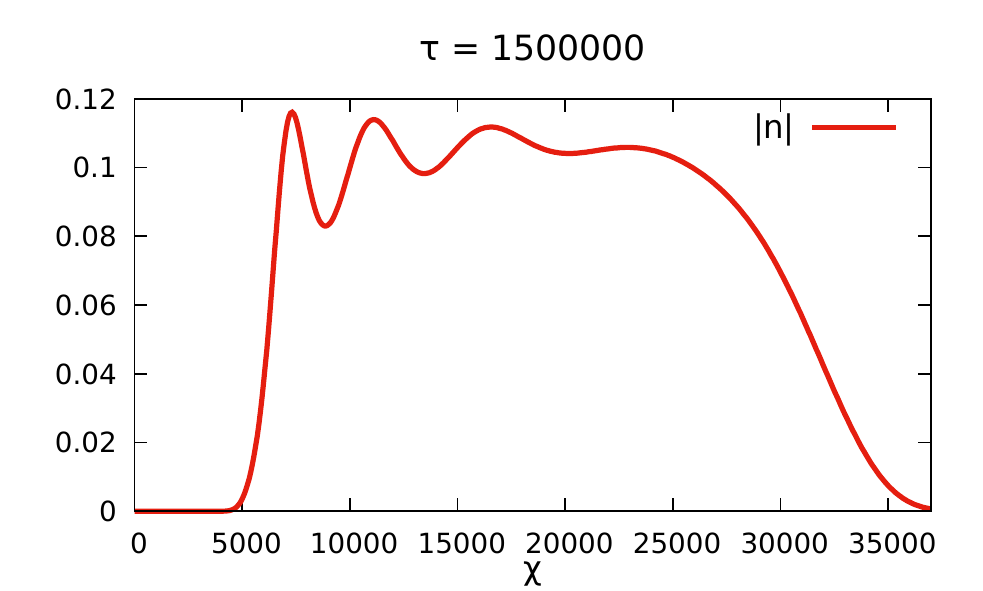}  
\caption{\footnotesize Spatial evolution of radiation (red), wiggler (green) and perturbed density (blue). Here, the initial wiggler maximum is $a_w (0) = 10^{-5}$, the maximum radiation starts with  a  $5 \%$ of wiggler, while the perturbed density starts from zero. The mean parameters of the system are: $n_b = 5.0\times 10^{18}$ cm$^{-3}$, $\gamma_e = 5.0$, $\lambda_w = 1000 $nm, and $\lambda_s \approx 10$ nm.}
\label{fig:nlinear}
\normalsize
\end{figure}

The full solution is represented in  Figure  \ref{fig:nlinear}, where the radiation, after stretching, is achieved by the wiggler,  starting an exchange of energy between them.  We note that the seed pulse slices into a leading pulse and a secondary train of pulses that gain energy at the expense of the wiggler, such that their maximum values are located near where $a_w$ is totally depleted.  In addition, we observe that the half-width of the leading pulse decreases with time.
With respect to the density perturbation, we note that the peaks coincide in number and position with the seed pulses. These results are closely related to the solutions obtained in laser amplification using stimulated Raman backscaterring in a plasma \cite{prlmsf1999}, where the behaviour of the three modes, pump, seed and plasmon, is described by the sine-Gordon equation, in which the asymptotic oscillations around $\pi$ value generates the pulse train in the seed radiation, which known as ``$\pi$-pulse''\cite{plagll1969}. This solution leads to amplification and compression of the leading pulse in a rate proportional to the time $\tau$. 

In order to show the relationship of Eqs. (\ref{as})-(\ref{n}) with the famous ``$\pi$-pulse" solution we assume that the perturbed density and wiggler have a much greater spatial variation than temporal one ($\partial_\chi \gg \partial_\tau$). By taking into account the parameters used in this work, the  quantum effects represented by $\beta_q$ are negligible in Eq. (\ref{n}) in comparison to the slippage $\beta_s - \beta_e$. Redefining the envelope amplitudes as,
$a_w = \sqrt{\frac{|C_w|}{|C_s| (\beta_s +\beta_w)}} \cos \left(\frac{\phi}{2} \right)$, $ n = \sqrt{\frac{|C_l|}{|C_s| (\beta_s -\beta_e)}} \sin \left(\frac{\phi}{2} \right)$, and  
$ a_s = -\frac{1}{2} \sqrt{\frac{|C_w||C_l|}{ (\beta_s +\beta_w)(\beta_s - \beta_e)}}  \frac{\partial \phi}{\partial \chi}$,
with $\phi (\chi,\tau)$ being the Bloch angle and substituting these expressions into Eq. (\ref{as}), we obtain the following sine-Gordon equation,
\begin{equation}
\frac{\partial^2 \phi}{\partial \chi \partial \tau} = \sin \phi,
\end{equation} 
whose the results are in agreement with numerical solutions described in Figure (\ref{fig:nlinear}).
According to the criteria established in Ref.\cite{spranglereview}, the FEL with the parameters defined above operates in the Raman regime, in which the transversal motion of the electrons, in the beam, is considered negligible with respect to longitudinal motion ($\gamma \approx \gamma_e$).


In conclusion, starting from the quantum fluid model presented in previous works, we have solved in space and in time the three-wave interaction equations, which describe an optical pumped FEL operating in the Raman Regime.
In our spatio-temporal  solution it is shown that the scattered radiation is amplified and divided into sub-pulses while it absorbs energy from the optical wiggler trough the negative energy mode associated to the plasma beam mode.  We also show that the leading pulse gets more energy than the others while the longitudinal waist decreases and has a shape similar to a  ''$\pi$-pulse'' solution from the sine-Gordon equation.
We emphasize that the behaviour presented here for the scattered radiation suggest us that FEL pumped by an optical wiggler work as laser amplification(seed) and is well described by three-wave interaction model. 
This result clearly shows that stimulated radiation emitted by FELs can achieve ultra high intensities, as it happens in a laser amplification by stimulated Raman process.  The difference between this two processes is that FELs operate with the approximate linear dispersion relation  $k_s \approx 4\gamma_e^2 k_w $, while the other with $k_s \approx k_w$.

\begin{acknowledgments}
We would like to thank the financial support of CAPES Brazil, and of the European Programme IRSES.
\end{acknowledgments}

\end{document}